\documentclass[superscriptaddress,amsmath,amssymb]{revtex4}

\usepackage{graphicx}
\usepackage{dcolumn}
\usepackage{bm}
\usepackage{color}
\usepackage{xcolor}
\usepackage{amsfonts}
\usepackage{amssymb}
\usepackage{array}
\usepackage{times}
\usepackage{epsfig}
\usepackage{epstopdf}
\usepackage{setspace}
\usepackage{dcolumn}
\usepackage{amsmath}
\usepackage{latexsym}
\usepackage{float}

\begin{document}

\title{Energy and information propagation in a finite coupled bosonic heat bath  }
\author{Fernando Galve}
\author{Roberta Zambrini}
\affiliation{IFISC  (UIB-CSIC), Instituto de
F\'{\i}sica Interdisciplinar y Sistemas Complejos, UIB Campus, 07122 Palma
de Mallorca, Spain}
\date{\today}

%

\begin{abstract}
The finite coupled bosonic model of reservoir introduced by Vasile {\it et al.} \cite{vasile} 
to characterize non-Markovianity, is used to study the different dissipative 
behaviors of a harmonic oscillator coupled to it when it is in resonance, close to resonance or far detuned. We show that information and 
energy exchange between system and heat bath go hand in hand because phonons are the carriers of both: in resonance free propagation of excitations 
is achieved, and therefore pure dissipation, while when far detuned the system can only correlate with the first oscillator in the bath's chain, 
leading to almost unitary evolution. In the intermediate situation we show the penetration of correlations and the formation of oscillatory (dressed state) 
behavior, which lies at the root of non-Markovianity.

\end{abstract}

\maketitle
\section{Introduction}

The quest for a perfect control of quantum processes so as to harness efficiently their promised advantage \cite{nielsen} over
the classical counterparts is leading to ever more perfect experiments and theoretical descriptions.
Initially the focus has been on the unitary control of systems whereby dissipation and decoherence were
just a drawback which had to be eliminated, but more recently there has been a shift of paradigm.  
Popular approaches to the description of dissipation are master equations for the 
system's density operators or Langevin equations for its operators \cite{dissip}. Typically, strong simplifications and
approximations were made in order to derive them, such as considering infinite reservoirs with zero memory effects, weak system-bath
couplings, etc. However there have been many developments with respect to that simple paradigm.  One direction is reservoir engineering\cite{disseng} 
where engineered dissipative processes are used to drive the system into a desired state or phase with interesting or
useful properties, like entanglement. Another direction is to revisit and broaden the description of decoherence and dissipation in 
presence of non-Markovian evolution and memory effects. This latter approach has seen a  
revival of interest due to recent works reporting possible advantages due to memory effects \cite{NMadv} with respect to Markovian dissipation and has been recently developed
with the introduction of more specific quantifiers \cite{nonmark1,nonmark2,nonmark3,nonmark4,nonmark5,nonmark6} of non-Markovianity.
These quantifiers are related to different features related to memory effects, including, for instance,
the divisibility of quantum evolutions or the flow-back of information from the environment into
the systems, and go beyond definitions well-known in classical systems \cite{VacchiniSmirne}  or to the 
Born-Markov approximation in the derivation of master equations \cite{dissip} (whereby the bath is assumed not to evolve and to react only to the present
state of the system and thus be memoryless). 

Many paradigmatic simplified models of dissipation have been revisited to study their  quantum non-Markovianity (see Refs. \cite{NMreview1,NMreview2} 
for an overview). In a former work we studied \cite{vasile} the
particular case of a finite 1D coupled bosonic model, also known as Rubin model\cite{rubin}, leading to Ohmic spectral density. Bosonic
chain models have been considered in the literature \cite{Ford} as interesting for the transmission of entanglement in quantum computing architectures \cite{pleniochain,harmochain},
or its creation when acting as a common heat reservoir \cite{lutz1,lutz2},
while more complicated structured bosonic models have been introduced as possible playgrounds for noiseless
subsystem creation and synchronization engineering \cite{synchronet}. The simple model of finite coupled reservoir \cite{vasile} allows to study
non-Markovian dissipation in full detail and to independently assess each factor contributing to memory effects in the dynamics. In addition this 
model reduces to the prototypical Ohmic spectrum for given values of the parameters in the thermodynamic limit, which is part of its beauty.
Furthermore, a bosonic bath producing non-Markovian effects can be turned into an interacting chain which itself dissipates into an Ohmic bath \cite{burghardt}, and in
general a star model of bosonic bath can always be turned into a coupled chain with given frequencies and nearest neighbor couplings \cite{prior}.

Since we are going to delve deeper in a particular aspect studied in Vasile {\it et al.} \cite{vasile} of the origin of non-Markovianity, namely
the system-bath resonant conditions, we will use the same model in its dimer configuration. There the authors argued \cite{vasile} that non-Markovian 
evolution came from close to resonance condition between system and bath. Purely resonant led to Markovian dissipation while purely nonresonant led to unitary
evolution, as is expected when bath and system are far detuned. In this work we will show that indeed in this kind of model energy and
information propagate from system to bath in the same way under the same regimes of resonance/detuning. We will show in Sect.3 that when close to, but not full, 
resonance is achieved information and energy flow back and forth from the system and `bound states' (oscillatory behavior) is formed between the system
and the first oscillators in the chain. In that regime we will also see that propagation has an evanescent wave like shape for both quantities. Finally, as
a side remark, we highlight the different propagation speeds of acoustic and optical phonons in the chain.

\section{The model}
The so called Rubin model consists of a harmonic oscillator, with canonical variables $\{q_S,p_S\}$ (the system), coupled to the first oscillator
of a harmonically coupled chain with strength $k$ (the heat bath). The chain is translation invariant and has open boundary conditions. It consists
of $N$ oscillators with onsite harmonic potential of frequency $\omega_0$ and harmonic nearest neighbor coupling. The introduction
of alternate couplings between bath oscillators allows for the appearance of a bandgap in the chain eigenfrequencies (this is interesting by itself,
since it can e.g. help induce topologically protected states in the chain \cite{schomerus}). 

\begin{figure*}[h!]
\includegraphics[width=\columnwidth]{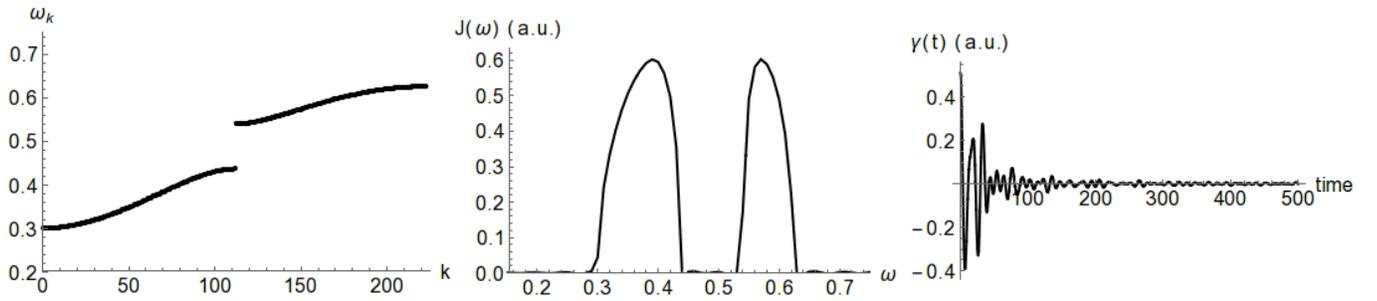}
\caption{{\it Left:} eigenfrequencies of the finite coupled chain with $\omega_0$ and couplings $g=0.1$, $h=0.05$. With these values the frequency ranges are from $\omega_0$ to $\sim\sqrt{\omega_0^2+2h}=0.436$ and from $\sim\sqrt{\omega_0^2+2g}=0.54$ to $\sim\sqrt{\omega_0^2+2(g+h)}=0.625$. {\it Middle:} spectral density $J(\omega)$ in arbitrary units. {\it Right:} damping kernel $\gamma(t)$. A purely Ohmic reservoir with infinite frequency cutoff would lead to a Dirac delta and therefore a memoryless damping (see text).}
\label{fig0}
\end{figure*}

The total Hamiltonian is then $H=H_S+H_B+H_{SB}$, with
\begin{eqnarray}
H_S&=&(p_S^2+\omega_S^2q_S^2)/2\nonumber\\
H_B&=&\sum_i(p_i^2+\omega_0^2q_i^2)/2+g\sum_i^{\rm{odd}}(q_{i+1}-q_i)^2/2+\nonumber\\
&&+h\sum_i^{\rm{even}}(q_{i+1}-q_i)^2/2\\
H_{SB}&=&-\kappa q_Sq_1\nonumber
\end{eqnarray}
(corresponding to the middle/red configuration in Fig. 1 of Ref. \cite{vasile})\footnote{It must be noted that in order to preserve translation invariance
of the chain, we need to impose extra onsite potential to the first and last oscillators, since they receive one onsite correction less from their interactions than 
the other neighbors)}.
The chain is initialized in the thermal Gibbs state which corresponds to $H_B$, while the system is initialized in a squeezed vacuum state.
The eigenfrequencies of the chain are shown in Fig.~ \ref{fig0}(Left) with its corresponding gap. The eigenfrequencies are similar to those of a closed dimer chain
$\omega_k^2=\omega_0^2+g+h+\sqrt{g^2+h^2+2gh\cos(2\pi k/N)}$ even if we are taking an open dimer chain, because our number of particles $N=225$ is big enough. The gap goes from the highest acoustic frequency $\sim\sqrt{\omega_0^2+2h}$ to the lowest optical frequency $\sim\sqrt{\omega_0^2+2g}$\cite{mermin}.
Dissipation caused by this bath leads to a Langevin equation for the system position observables
\begin{equation}
\ddot{q}_S+\tilde{\omega}_S^2q_S+\int_0^t ds \gamma(t-s)\dot{q}_S(s)=\xi(t)
\end{equation}
where $\tilde{\omega}$ is the system frequency renormalized through its interaction with the bath (we will take it to be $\omega_S$ since we 
will work in relatively weak coupling, $\kappa=10^{-4}$), $\xi(t)$ is a Langevin forcing \cite{vasile} and the damping kernel is
\begin{equation}
\gamma(t)=\sum_i \frac{g_i^2}{\omega_i^2}\cos(\omega_it)
\end{equation}
(with $g_i$ the coupling strength of system and chain's eigenmode $i$). The latter can be seen in fig~\ref{fig0} right. 
The spectral density quantifies the strength with which the system couples to a given part of the bath spectrum, as defined by
\begin{equation}
J(\omega)=\sum_i \frac{g_i^2}{\omega_i}\delta(\omega-\omega_i)
\end{equation}
but can also be obtained from the damping kernel through
\begin{equation}
J(\omega)=\omega\int_0^\infty dt \gamma(t)\cos(\omega t).
\end{equation}
It is shown in fig~\ref{fig0} middle. A purely Ohmic reservoir has a spectral density $J(\omega)\propto\omega$ with cutoff frequency at infinity. It is easy to show that this corresponds to a Dirac delta function for $\gamma(t)$ and therefore a time independent damping for the system oscillator (thus memoryless).
In our case, since the highest bath frequency ($\omega_c$) is finite, the damping kernel `remembers' a finite time evolution of the system ($\sim2\pi/\omega_c$).
Were we to set $g=h$ and $\omega_0=0$ we would have the typical Ohmic model with cutoff frequency $\sqrt{\omega_0^2+4g}$.

\begin{figure*}[h!]
\includegraphics[width=17cm]{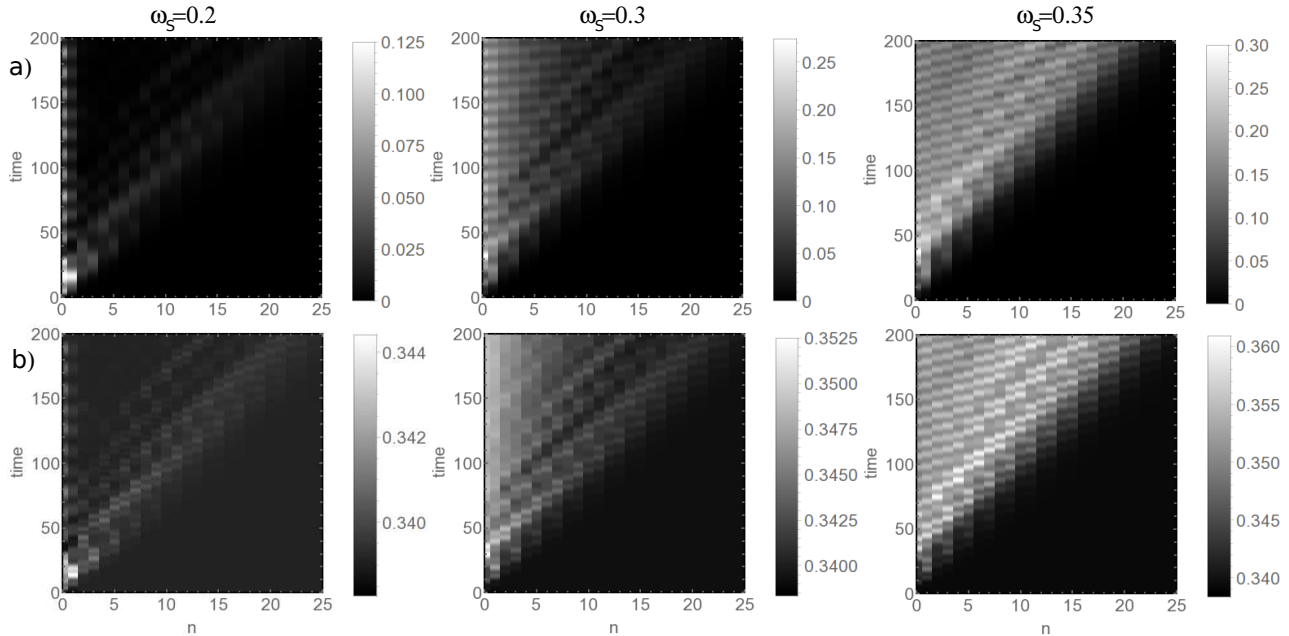}
\caption{Time evolution of mutual information between the system and the $n$-th bath oscillator (a), and energy of the $n$-th oscillator (b). The three cases are $\omega_S=0.2$ (nonresonant),$0.3$ (edge, almost resonant) and $0.35$ (resonant). Only the resonant case propagates freely along the chain. The detuned case (left) only exchanges excitation between system and the first oscillators in the chain. The edge case forms a dressed state between system and a set of close-by bath oscillators. The energy of the system is not shown.}
\label{fig1}
\end{figure*}

\section{Resonance conditions for information and energy exchange}

In Vasile {\it et al.}\cite{vasile} we argued that the flow between the system and bath has three different regimes, both for energy and for information. 
When they are in resonance, energy is dissipated from the system with no flow back. Also information is lost, entropy of the system grows almost monotonically as the bath `learns' more 
and more about the system and gets entangled with it. In the opposite regime, when far detuned, they practically do not interact and thus the system evolves almost unitarily, without 
dissipation or decoherence. However, in the nearby of frequency edges of the reservoir, energy can flow to the bath but not in a completely resonant fashion, and thus partially flows 
back intermittently, causing oscillatory behavior both for system observables and for its decoherence properties. This was conjectured to be the root of non-Markovian behavior
as actually, near to edges of the spectrum gap, the system showed the largest non-Markovianity with two different quantifiers, 
a measure and a witness (see Fig.~3 in Ref. \cite{vasile}).

Here we will show that indeed these resonance conditions are the origin of these three regimes, and that information behaves exactly in the same way as energy. 
In order to quantify information flow from system to bath, we will use the quantum mutual information between system and each of the oscillators in the reservoir chain. 
It is defined as
\begin{equation}
I(S:n)=S(\rho_S)+S(\rho_n)-S(\rho_{S,n})
\end{equation}
with $\rho_S$ the reduced density operator of the system, $\rho_n$ that of the $n$-th oscillator in the chain and $\rho_{S,n}$ the joint density matrix. 
It quantifies the maximum amount of information that either $n$ or $S$ has about the other oscillator. 

\begin{figure*}[h!]
\includegraphics[width=17cm]{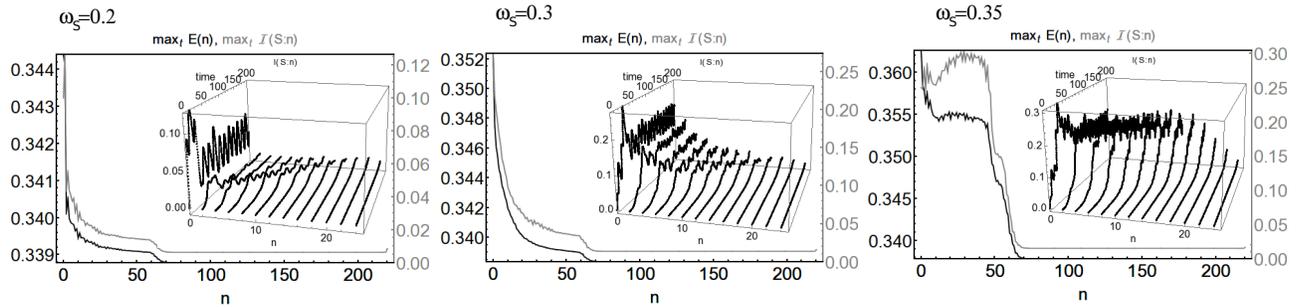} 
\caption{Maximum value of the energy $E(n)$ of bath oscillator number $n$ (black) and of mutual information between system and $n$-th oscillator (gray), along a time period $t\in[0,500]$. Insets show the time evolution of only mutual information. From left to right the system has frequencies $\omega_S=0.2$ (nonresonant),$0.3$ (edge, almost resonant) and $0.35$ (resonant).
}
\label{fig2}
\end{figure*}

In Fig.~\ref{fig1} we can see the time evolution of energy and mutual information in the three regimes $\omega_S=0.2,0.3,0.35$, i.e. nonresonant, edge and resonant. 
It is evident that energy and mutual information propagate very similarly, as expected from the fact that both are carried through phonons in the chain.
In the nonresonant case $\omega_S=0.2$ the excitations try to enter the chain but are not allowed because they cannot resonate with eigenmodes, i.e. phonons cannot carry energy 
of that frequency. This is the effect of a band-gap, well-known in solid state \cite{semic} and in photonic crystals \cite{PC}, and here discussed focusing on the 
microscopic  energy and information flows. 
Therefore the system oscillator gets correlated {\it only} with the first element of the chain and only a tiny bit (temporarily) with the rest. This can also be seen in the insets of Fig.~\ref{fig2}.

In contrast, under resonance conditions ($\omega_S=0.35$) for the system and the phonon frequencies,
the energy and correlations propagate freely along the chain as shown in Fig.~\ref{fig1} (Right) where the front of excitation/correlation is clearly recognized, linearly advancing in time. 
Indeed, in contrast with the previous case, here the system loses energy and thermalizes on a time scale $\propto \gamma^{-1}$.

In the intermediate case (when the system frequency is at the edge of the band, $\omega_S=0.3$ ) correlations are able to enter partially into the chain, 
but with an evanescent wave profile, i.e. the correlations decay exponentially fast with the distance. In contrast to the detuned case, these correlations {\it survive} for long times, 
which can be understood as the formation of bound (dressed states) between the system and the first oscillators of the chain \cite{quang}.

\begin{figure*}[h!]
\includegraphics[width=8cm]{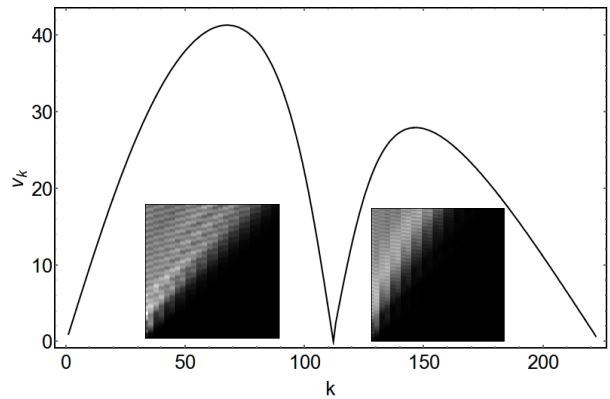} 
\caption{Group velocity of eigenmodes (phonon velocities) in the chain. The highest speed of the acoustic band is higher than that of the optical band. Insets show the propagation of mutual information for the cases when the system is resonant with each band, i.e. $\omega_S=0.35$ (left) and $\omega_S=0.57$ (right), clearly confirming the phonon propagation mechanism for the exchange of energy and correlations.}
\label{fig3}
\end{figure*}

But how far can correlations and energy penetrate the chain? This is shown most clearly in Fig.~\ref{fig2}, where we have plotted the maximum value of both 
along a time period of $t=500$, also representing the detailed dynamical evolution of the relevant bath oscillators. The closer we are to resonance, the more
inwards they reach into the chain. In case of resonance, the reach the longest distance that the fastest excitations can carry them, i.e. the light cone of phonons.
It is interesting to note in this sense, that the propagation velocity of the two different bands is different. This is shown in Fig.~ \ref{fig3}, where we 
represented the phonon velocities $v_k=d\omega_k/dk$. It is interesting that the low-frequency (acoustic) band propagates faster than the optical band, which
is clearly seen when we plot the mutual information between system and bath oscillators when the system is resonant with the acoustic ($\omega_S=0.35$, left plot) 
or the optical ($\omega_S=0.57$, right plot) band.

\section{Conclusions}
The availability of a finite coupled model of reservoir which reproduces the typical Ohmic spectral density or can be tuned to reproduce structured baths (as for the dimer here considered)
has shown to be a great advantage. Indeed, the possibility to go from the picture of real oscillators in the coupled chain to the picture of phononic excitations propagating at given velocities along the chain, has served to fully understand the exchange of information and energy between system and bath. From this understanding it is easy to  capture the important
regimes where non-Markovian evolution can arise, precisely when the system is close to resonant but not completely. This leads to the back and forth flow of information and to the formation of bound/dressed states, meaning that the coherence of the system oscillates and therefore that the measures of non-Markovianity will yield finite values.
Also, the fact that there is an underlying microscopic model of reservoir allows us to change its parameters and create a gap. This gap not only adds spectral regions where the system can form dressed oscillatory states but also divides the bath's frequency range into acoustic and optical bands. Remarkably, the high frequency (optical) band is seen to propagate slowlier than the acoustic one.

In summary we have shown that the microscopic model for dissipation introduced in Ref. \cite{vasile} can explain when non-Markovian behavior will appear mostly by appealing to resonance conditions and to the understanding of the flow of excitations into the bath. The most non-Markovian regime is indeed shown to be related to the oscillatory behavior of observables when a dressed state is formed between the system and a small set of close-by oscillators in the bath.

\section*{Acknowledgments}
We would like to acknowledge funding from  MICINN, MINECO, CSIC, EU
commission, FEDER funding under Grants No. FIS2007-60327 (FISICOS) 
and No. FIS2011-23526 (TIQS), postdoctoral JAE program (ESF), and COST Action MP1209.

\vspace*{-5pt}   

\end{document}